\documentclass[twocolumn,aps,prb,superscriptaddress,notitlepage]{revtex4}
\usepackage{lmodern} 
\usepackage[T1]{fontenc}
\usepackage{textcomp}
\usepackage[scr=boondoxo,frak=boondox,bb=boondox]{mathalfa}
\usepackage{graphics,epsfig,amsfonts,amssymb,amsmath,color}
\usepackage{subfig}
\usepackage{footnote}
\newcommand{\beginsupplement}{%
        \setcounter{table}{0}
        \renewcommand{\thetable}{S\arabic{table}}%
        \setcounter{figure}{0}
        \renewcommand{\thefigure}{S\arabic{figure}}%
        \setcounter{equation}{0}
        \renewcommand{\theequation}{S\arabic{equation}}%
     }

\begin{document}
\title{Entanglement control and magic angles for acceptor qubits in Si}
\author{J. C. Abadillo-Uriel}
\affiliation{Materials Science Factory, Instituto de Ciencia de Materiales de Madrid, ICMM-CSIC, Cantoblanco, E-28049 Madrid (Spain).}
\affiliation{School of Physics and Centre for Quantum Computation and Communication Technology, The University of New South Wales, Sydney 2052, Australia}
\author{Joe Salfi}
\affiliation{School of Physics and Centre for Quantum Computation and Communication Technology, The University of New South Wales, Sydney 2052, Australia}
\author{Xuedong Hu}
\affiliation{School of Physics and Centre for Quantum Computation and Communication Technology, The University of New South Wales, Sydney 2052, Australia}
\affiliation{Department of Physics, University at Buffalo, SUNY, Buffalo, New York 14260-1500, USA}
\author{Sven Rogge}
\affiliation{School of Physics and Centre for Quantum Computation and Communication Technology, The University of New South Wales, Sydney 2052, Australia}
\author{M. J. Calder\'on}
\affiliation{Materials Science Factory, Instituto de Ciencia de Materiales de Madrid, ICMM-CSIC, Cantoblanco, E-28049 Madrid (Spain).}
\author{Dimitrie Culcer}
\affiliation{School of Physics and Australian Research Council Centre of Excellence in Low-Energy Electronics Technologies, UNSW Node, The University of New South Wales, Sydney 2052, Australia}
\date{\today}
\begin{abstract}
\noindent\textbf{ABSTRACT:} Full electrical control of quantum bits could enable fast, low-power, scalable quantum computation. Although electric dipoles are highly attractive to couple spin qubits electrically over long distances, mechanisms identified to control two-qubit couplings do not permit single-qubit operations while two-qubit couplings are off. Here we identify a mechanism to modulate electrical coupling of spin qubits that overcomes this drawback for hole spin qubits in acceptors,that is based on the electrical tuning of the direction of the spin-dependent electric dipole by a gate. In this way, inter-qubit coupling can be turned off electrically by tuning to a ``magic angle'' of vanishing electric dipole-dipole interactions, while retaining the ability to manipulate the individual qubits.
This effect stems from the interplay of the $\rm T_d$ symmetry of the acceptor state in the Si lattice with the magnetic field orientation, and the spin-3/2 characteristic of hole systems. Magnetic field direction also allows to greatly suppress spin relaxation by phonons that limit single qubit performance, while retaining sweet spots where the qubits are insensitive to charge noise. Our findings can be directly applied to state-of-the-art acceptor based architectures, for which we propose suitable protocols to practically achieve full electrical tunability of entanglement and the realization of a decoherence-free subspace.

\vspace{0.2cm}\noindent\textbf{KEYWORDS:} \emph{Qubit, holes, electrical tuning, entanglement, sweet spot, decoherence free subspace}
\end{abstract}
\maketitle

A scalable quantum computer architecture requires long qubit coherence times~\citep{NielsenChuang, FowlerPRA2012} and fast high-fidelity control of single-qubit and two-qubit operations. For solid-state spin qubits~\citep{LossPRA1998, KaneNature1998, PettaScience2005}, silicon offers improved spin lifetimes~\citep{TyryshkinJOP2006, MuhonenNatNano2014, TahanPRB2005, WangPRB2010, RaithPRB2011, VeldhorstNatNano2014}, elimination of nuclear spin induced decoherence by isotope purification~\citep{ItohPRB2004, ItohMRS2014}, absence of spin relaxation by piezoelectric phonons~\citep{EhrenreichPR1956}, and compatibility with Si microtechnology.~\citep{MaurandNatCom2016} Coupling of spin qubits to electric fields has recently attracted much attention to improve qubit manipulation rates~\citep{LossPRL2007, NowackScience2007, MedfordPRL2013, SchirmNatureNano2013, RomhanyiPRB2015}, while electric fields are significantly easier to apply and localize than magnetic fields, and use much less power.~\citep{MortonNature2011} Coupling to electric fields also open new possibilities for two qubit gates mediated by electric fields\citep{FlindtPRL2006, GolovachPRB2006} or microwave photons.\citep{TrifPRB2008}

\begin{figure}
\leavevmode
\includegraphics[clip,width=0.44\textwidth]{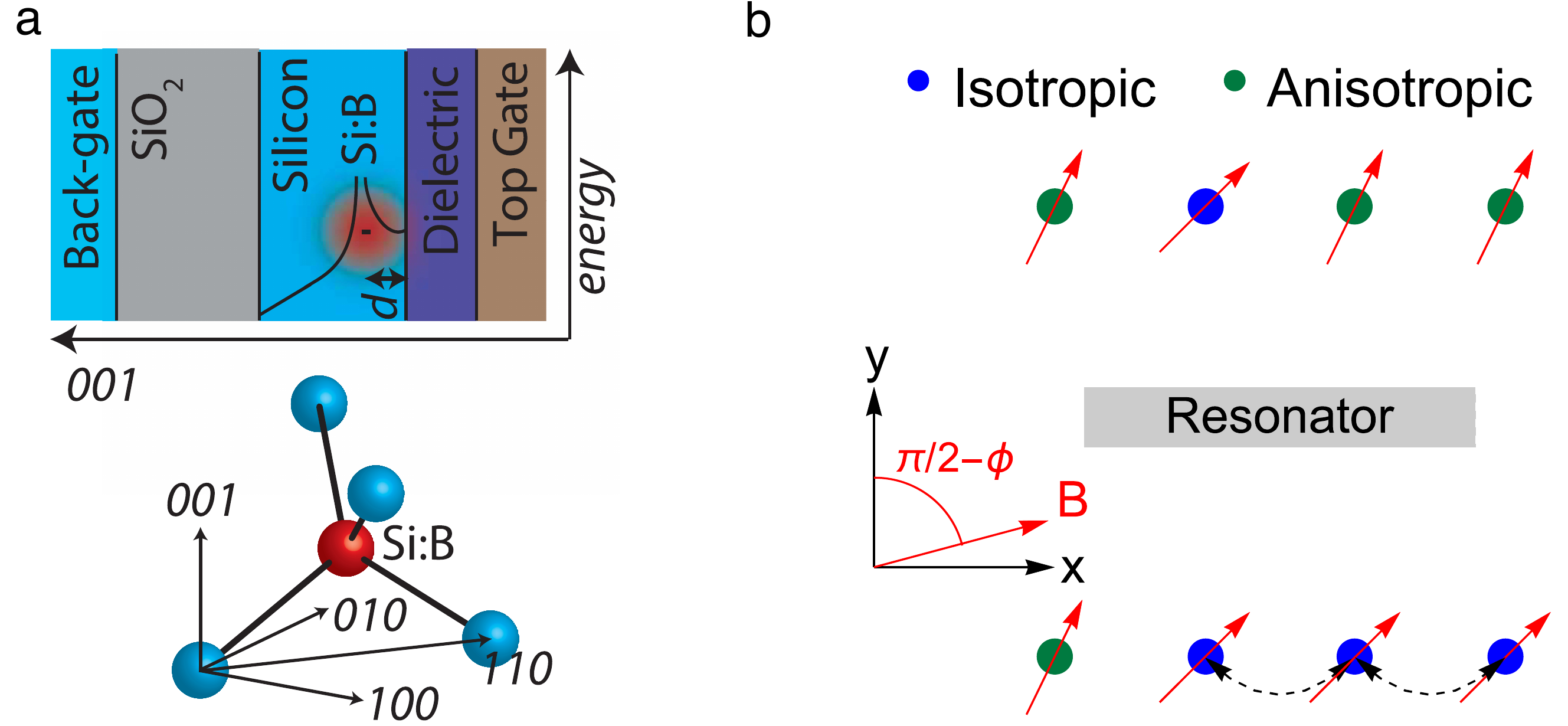}
\caption{(a) Top: Layered heterostructure in the (001) direction. $d$ denotes the depth of the acceptor beneath the top gate interface. Bottom: representation of the local symmetry of the acceptor in the Si lattice. (b) Schematic view of the nearest-neighbor two-qubit interactions for Protocol 2 with $\phi=15^\circ$. The orientation of the magnetic field within the plane is given by $\phi$, the angle with respect to the (100) direction. The top gate tunes the qubits between the different configurations with different charge-dipole orientations (red arrows) used to couple or decouple neighboring qubits.}
\label{fig:sketch}
\end{figure}

Electrical spin manipulation can be achieved with holes in Si~\cite{Dykmanpreprint} thanks to the intrinsically large spin orbit interaction (SOI) in the Si valence band.~\citep{SzumniakPRL2012, PalyiPRL2012, FriesenPRL2004, SalfiPRL2016, SalfiNano2016, WinklerPRB2004, CulcerPRL2006, WinklerSST2008, Winklerbook} Recently, an acceptor-based quantum information processing platform was introduced~\cite{SalfiPRL2016}, in which inversion symmetry breaking by the interface (Fig. \ref{fig:sketch}(a)) gives rise to a Rashba interaction that couples the spin to in-plane electric fields, enabling fast electrical spin manipulation via electric dipole spin resonance (EDSR). Importantly, sensitivity to charge noise which produces dephasing is suppressed to first order. Yet, two limitations of the conventional electrically driven spin qubits remain: (i) $T_1$ controlled by phonons cannot be significantly enhanced without sacrificing qubit manipulation rates, and (ii) inter-qubit couplings based on electric dipole-dipole interactions can only be turned off by deactivating one of the qubits.

Here we show that these limitations can be lifted by exploiting an unconventional interaction between the acceptor bound hole and the in-plane magnetic field direction that derives from the $T_d$ symmetry of the acceptor in the Si lattice. The $T_d$ term's strength and dependence on the top gate electric field allows the qubit's spin polarization to be controlled by the gate rather than the magnetic field, so that magnetic field orientation allows an unusually large tuning of qubit couplings to the environment. A decoherence free subspace (DFS) is possible where the qubit is essentially decoupled from \textit{phonons}, while further improving previously identified insensitivity to \textit{charge noise}.\cite{SalfiPRL2016}  Moreover, the top gate can be used to turn on and off the electric dipole-dipole coupling while allowing both qubits to be manipulated independently when the coupling is off.  The underlying mechanism is a rapid electrical control of the charge dipole orientation of the individual qubits.

We analyze this new two-qubit coupling mechanism and propose two protocols that allow fast, independent, fully electrical single-qubit and long-distance two-qubit manipulations that are compatible with long coherence times. The new two-qubit coupling mechanism and enhanced spin lifetimes predicted here greatly improve the prospects for a practical implementation of quantum information using extensively studied hole spin systems.~\citep{KatsarosNatNano2010, WatzomgerNanoLett2016, NenashevPRB2003, MalissaAPL2004, AresPRL2013, SrinivasanPRB2016, PradoPRB2004, KuglerPRB2009, AndlauerPRB2009, BennettNatComm2013} For hole spins bound to acceptor atoms many milestones have already been achieved experimentally, from the placement of acceptors near an interface~\citep{MolAPL2015}, to the measurement of single-acceptor states~\citep{vdHeijden2014}, and coupling between two acceptors.~\citep{SalfiNatComm2016,vdHeijden2017} The results reported here also highlight the advantages of acceptors versus quantum dots as a suitable platform for hole spin qubits.

Figure~\ref{fig:sketch}(a) shows the layered geometry in the vertical direction of the heterostructure for a single acceptor as a qubit, while Fig.~\ref{fig:sketch}(b) shows schematically the coupling between neighboring qubits in a 2D array under one of our proposed protocols.  The theoretical description of a single acceptor relies on the Hamiltonian (see Supp. Mat. \citep{suppl-material})
\begin{equation}
H=H_{KL}+H_{BP}+H_c+H_{\text{inter}}+H_F+H_B+H_{\rm T_d} \,
\end{equation}
which contains the details of the Si valence band via the Kohn-Luttinger Hamiltonian~\cite{KohnPR1955} $H_{KL}$ including cubic symmetry terms, the strain Bir-Pikus term~\cite{BirJPCS1963,BirJPCS1963II} $H_{BP}$, the acceptor Coulomb potential $H_c$, the $(001)$ Si/SiO$_2$ interface $H_{\text{inter}}$, the interaction with electric field $H_F$, and magnetic field $H_B = g_1 \mu_B \mathbf{B} \cdot \mathbf{J} + g_2 \mu_B \mathbf{B} \cdot \mathbf{J^3}$, where $\mathbf{J}$ represent the spin-3/2 matrices. The projection of the spin-3/2 onto the axis perpendicular to the interface, $\hat{\bm z}$, is $m_J=\pm 3/2$ for the \textit{heavy holes} (HH) and $m_J=\pm 1/2$ for the \textit{light holes} (LH). Tensile strain gives the qubit a LH character~\cite{SalfiPRL2016, Abadillo-UrielNJP2017}, ensuring a strong Zeeman interaction with an \textit{in-plane} magnetic field.~\cite{note1}

\begin{figure}[t]
{\includegraphics[clip,width=0.5\textwidth]{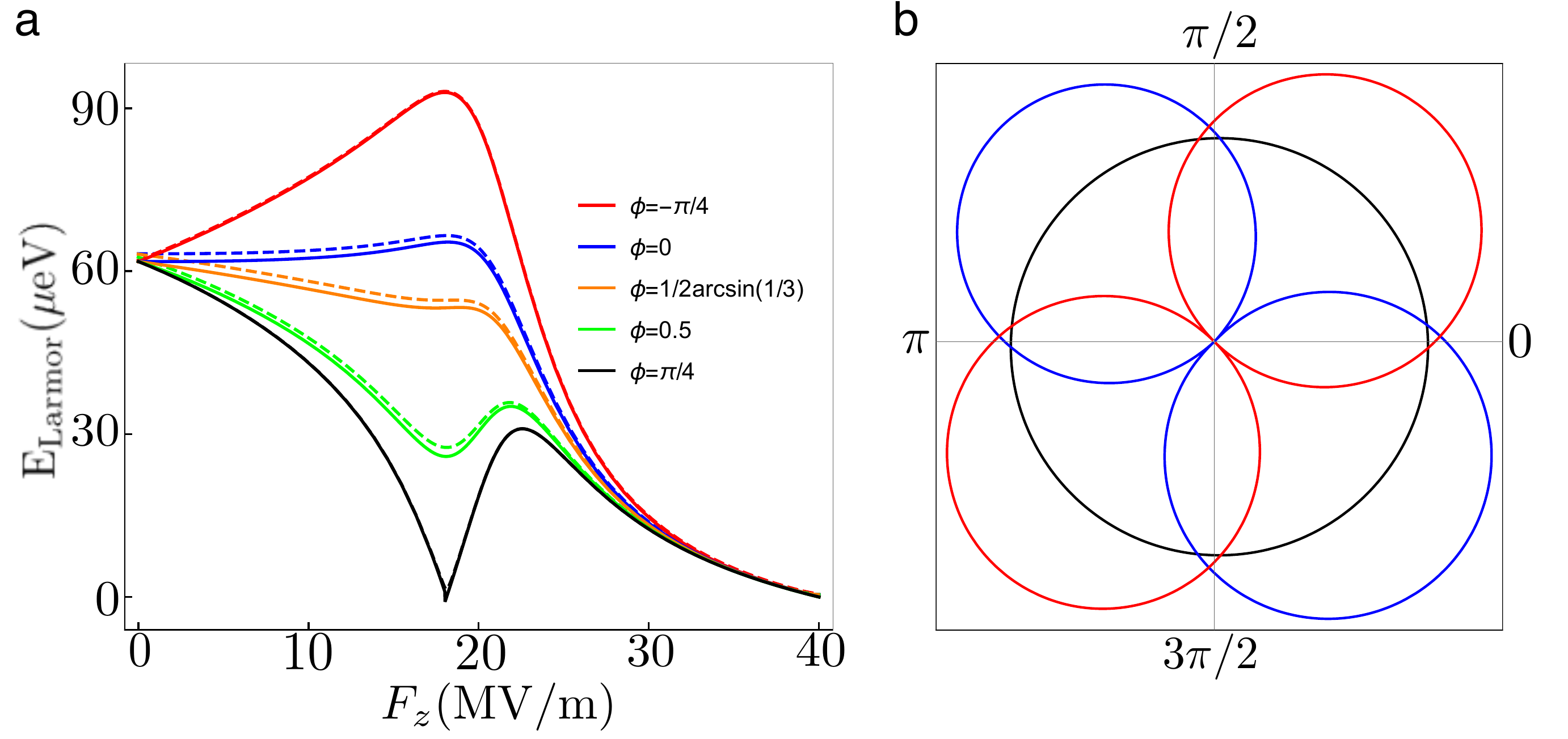}}
\caption{(a) Larmor energy for different magnetic field orientations as a function of the vertical electric field with $B=0.5$ T and $d=4.6$ nm. The Larmor energy at the isotropic sweet spot ($F_z=18.1$ MV/m) is enhanced for $\pi/2<\phi<\pi$ and depressed for $0<\phi<\pi/2$ with a minimum at $\phi=\pi/4$. At this point the two lowest levels cross. The dashed lines correspond to the numerical results. When $\phi=1/2\arcsin(1/3)$ there is a single sweet spot to second order. (b) \textit{g}-factors as a function of the magnetic field orientation of the qubit states at $F_z=0$ (black), isotropic sweet spot $F_z=18.1$ MV/m (blue) and the upper branch at $F_z=22.5$ MV/m (red). The latter corresponds to the upper branch composition $a_L=\sqrt{3}/2$ and $a_H=1/2$.
}
\label{fig:larmor}
\end{figure}

The tetrahedral symmetry of the acceptors gives rise to a linear coupling to the electric field of the form $H_{\rm T_d}=\frac{p}{\sqrt{3}}\left(\{J_y,J_z\}{F_x}+\{J_x,J_z\}{F_y}+\{J_x,J_y\}{F_z}\right)$, where the \textit{quadrupolar} terms involving products of two spin matrices have no counterpart in spin-1/2 electron systems (in which they are either the identity or zero). Here $p$ is an effective dipole moment that can be calculated~\cite{KopfPRL1992} by $p=e\int_0^a f^*(r)r f(r)$ with $a$ the lattice constant of the host material, and $f(r)$ the radial bound hole envelope function. For a boron acceptor in Si $p=0.26$ Debye.  This value is larger in deep acceptors, with a smaller Bohr radius.~\cite{KopfPRL1992} $H_{\rm T_d}$ is also linearly proportional to the strength of the electric field.  We neglect other allowed T$_d$ symmetric terms as their coupling constants are much smaller.~\citep{BirJPCS1963II}

We use the effective mass approach described in Ref. ~\onlinecite{AbadilloUrielNanotech2016}, however in this work we consider a magnetic field $\mathbf B$ with an arbitrary in-plane orientation characterised by an angle $\phi$. In the basis $\{3/2,1/2,-1/2,-3/2\}$, the effective Hamiltonian is~\cite{SalfiPRL2016,Abadillo-UrielNJP2017}
\begin{equation}
H_{\rm eff}=\begin{pmatrix}
0 & \frac{\sqrt{3}}{2}\varepsilon_Ze^{-i\phi} & -ipF_z & 0 \\
\frac{\sqrt{3}}{2}\varepsilon_Ze^{i\phi}  & \Delta_{HL} & \varepsilon_Ze^{-i\phi} & -ipF_z \\
ipF_z & \varepsilon_Ze^{i\phi} & \Delta_{HL} & \frac{\sqrt{3}}{2}\varepsilon_Ze^{-i\phi} \\
0 & ipF_z  & \frac{\sqrt{3}}{2}\varepsilon_Ze^{i\phi} & 0
\end{pmatrix},
\label{eq:H_LH}
\end{equation}
where $\phi=0$ for $\mathbf B$ along the $x$-direction, given by one of the main crystal axes. The cubic \textit{g}-factor $g_2$~ (ref \onlinecite{KopfPRL1992}) is not explicitly shown in Eq.~\ref{eq:H_LH}, but is included in the numerical calculations. The Zeeman term is $\varepsilon_Z=g_1 \mu_B B$. The HH energy is set to zero for $F_z=0$, and $\Delta_{HL}$ is the HH-LH energy difference. The qubit is defined by the two levels making up the spin-split ground state. At zero fields, the strain conditions are such that the lower (qubit) branch is of LH character ($\Delta_{HL}<0$) while the upper branch is of HH character. These LH and HH branches interact and they anticross at a particular value of the vertical field $F_z$ (ref \onlinecite{SalfiPRL2016}).

As Fig.~\ref{fig:sketch}(a) shows, the local tetrahedral symmetry of the acceptor makes a clear distinction between the main crystal axes and any other direction. It is represented by the term $H_{\rm T_d}$ in the Hamiltonian, which governs the qubit interaction with electric fields, and becomes more pronounced as the top gate voltage is increased, generating a mixing between HH and LH in the two branches. The interplay between the terms with tetrahedral symmetry $H_{\rm T_d}$ and the usual Zeeman interaction $H_B$ gives rise to a new and counterintuitive magnetic field orientation dependence of the qubit properties with no analog for spin-1/2 electrons. Fig.~\ref{fig:larmor} shows the strong dependence of the qubit frequency on the magnetic field orientation.

Adjusting the in-plane magnetic field direction influences the properties of previously identified sweet spots where the derivative of the qubit energy vanishes as a function of electric field. At $\phi=0$ one sweet spot is located at $F_z=0$ while the other resides at a finite value of $F_z$ (18.1 MV/m in Fig.~\ref{fig:larmor}(a) for $d=4.6$ nm), which depends on the acceptor depth. The sweet spot at $F_z=0$ moves in $F_z$ as a function of $\phi$ and is hence called the {\it anisotropic sweet spot}. On the other hand, the large field sweet spot, where LH and HH levels in the qubit mix with probability amplitudes $a_L=1/2$ and $a_H=-\frac{\sqrt{3}}{2}$ respectively, remains at the same value of $F_z=18.1$ MV/m, and is thus called the {\it isotropic sweet spot}. We find that for the particular values $\phi=1/2\arcsin(1/3)+n\pi$, and $\phi=1/2\left(\pi-\arcsin(1/3)\right)+n\pi$, the isotropic and anisotropic sweet spots fuse into a single sweet spot that makes the qubit insensitive to charge noise up to second order, see Fig~\ref{fig:larmor}(a). In this case the energy dispersion is flat at the sweet spot within a 2-3 MV/m window.

\begin{figure}
\includegraphics[clip,width=0.5\textwidth]{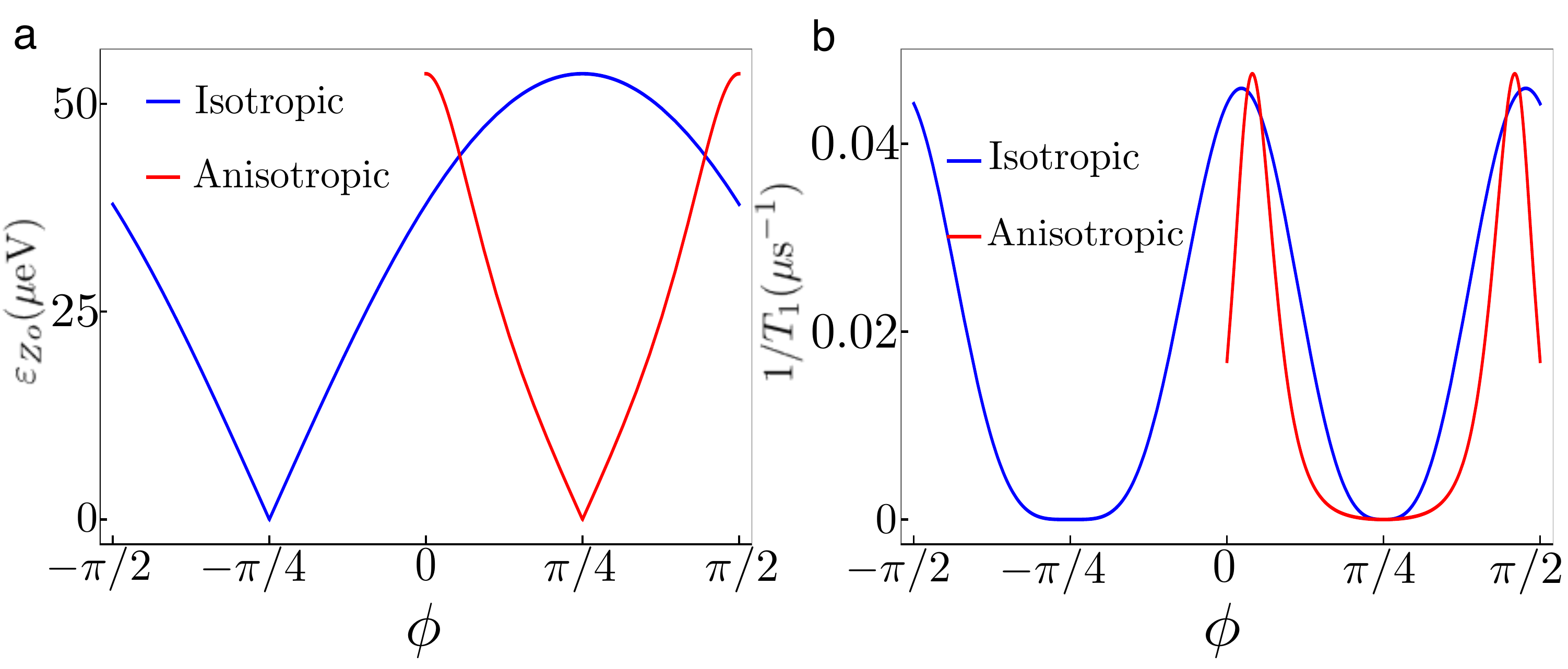}
\caption{(a) Value of the coupling between lower and upper branches $\varepsilon_{Zo}$ as a function of $\phi$. When this value is zero the qubit is in a DFS. (b) Inverse of the phonon-induced relaxation times $T_1$ with $B=0.5$T. The DFS is at $\phi=-\pi/4$ and $\phi=\pi/4$ for the isotropic and anisotropic sweet spots respectively. In the isotropic sweet spot the qubit becomes insensitive to the magnetic field when $\phi=\pi/4$.
}
\label{fig:relaxation}
\end{figure}

The spin lifetime enhancement and charge dipole orientation mechanism for controlling two-qubit coupling rely on gate voltage control of the qubit's spin polarization and appropriate choice of the magnetic field relative to the polarization.  Since spin polarization of a conventional spin 1/2 qubit only depends on the magnetic field, we first explain the mechanism to electrically vary spin polarization, and how the orientation of the effective spin polarization\citep{suppl-material} relative to magnetic field influences single qubit properties. We note that in our acceptor qubit, the HH and LH levels have different projections of total angular momenta.  Consequently, adjusting the LH-HH hybridization using the gate electric field $F_z$ influences the spin polarization.  The orientation of the magnetic field with respect to this spin polarization influences qubit dynamics including coupling to \textit{e.g.} phonons by changing the magnetic coupling $\varepsilon_{Zo}$ to the excited branch (see Fig.\ref{fig:relaxation}(a)).  For example, when the magnetic field is colinear to the spin polarization, the latter becomes a constant of the motion and the following holds

\begin{equation}
\left[H_{\rm inter}+H_{\rm T_d},H_B\right] = 0.
\label{eq:commutator}
\end{equation}
Here, the qubit completely decouples from the upper branch, and environmental effects on qubit dynamics are suppressed in a decoherence free subspace (DFS). This DFS differs from conventional ones, which arise from symmetries of the encoded states.\cite{note3}
Conversely, when the magnetic field points perpendicularly to the effective spin polarization of a particular branch the hole can no longer sense the magnetic field and the effective \textit{g}-factor becomes zero, as shown in Fig.~\ref{fig:larmor}(b).

Given that magnetic field orientation can suppress magnetic coupling $\varepsilon_{Zo}$ to the upper branch, and that acoustic phonons mix the qubit and upper branches, magnetic field orientation suppresses the spin relaxation due to acoustic phonons, see Fig.\ref{fig:relaxation}(b). The induced relaxation is proportional to the coupling $\varepsilon_{Zo}$ and inversely proportional to the qubit-upper branch energy difference $\Delta$ (ref \onlinecite{SalfiPRL2016}): $1/T_1\propto E_{\rm Larmor}(\phi)^3\left[\varepsilon_{Zo}(\phi)/\Delta\right]^2$. For an acceptor at 4.6 nm from the interface at the isotropic sweet spot, with $B=0.5$ T pointing along one of the crystal axes, $T_1 \approx 20$ $\mu$s. While $T_1$ can be enhanced trivially by suppressing the qubit frequency $\hbar\omega$, it is more interesting to enhance $T_1$ by suppressing the coupling $\varepsilon_{Zo}$, which can be quite effective since the $4\times4$ manifold is very well isolated from higher excited states (by $\sim 20$ meV).  Since the qubit coupling to in-plane electric fields providing single qubit gates by EDSR is $D\propto (\varepsilon_{Zo}(\phi)/2\Delta)$, the qubit operations are slowed down. However, $D$ decreases with $\varepsilon_{Zo}$ while $T_1$ is enhanced much faster with $1/\varepsilon_{Zo}^2$ (see \onlinecite{suppl-material}), so the number of single-qubit operations per qubit lifetime $r$ is enhanced even if the operations are slowed down.\citep{suppl-material} For instance, when $\phi=0$, $r=10^5$, and it can be enhanced to $10^6$ for $\phi=30^\circ$ or to $10^7$ for $\phi=40^\circ$.
\begin{figure*}
{\includegraphics[clip,width=1.\textwidth]{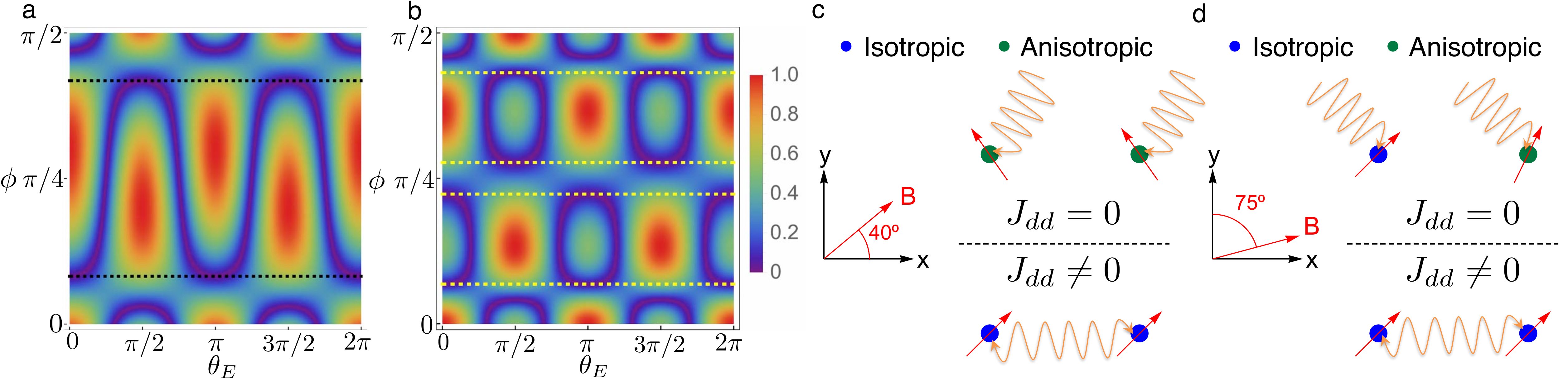}}
\vskip -0.2cm
\caption{Normalized function $G(F_z^a,F_z^b,\phi,\theta_E)$ that modulates the two-qubit coupling, see Eq.~\ref{eq:J}. In case (i) $G(F_z^a,F_z^b,\phi,\theta_E)=1$ independently of the relative orientation. (a) Case (ii): Qubit $a$ at the isotropic sweet spot while qubit $b$ at the anisotropic one. The black dashed lines indicate $\phi=15^{\circ}$ and $75^{\circ}$, orientations for which the coupling is suppressed in one main axis while enhanced in the other. (b) Case (iii): Both qubits at the anisotropic sweet spot. There are now four different magnetic field orientations ($\phi=12^{\circ}, 78^{\circ}, 40^{\circ}$ and $50^{\circ}$) leading to axis selective coupling suppression (yellow dashed lines). (c) Scheme of protocol 1. The anisotropic-anisotropic combination is used for single qubit operations. Two-qubit operations are activated in the isotropic-isotropic combination. (d) Scheme of protocol 2. The isotropic-anisotropic combination is used for single qubit operations. Two-qubit operations are activated in the isotropic-isotropic combination.}
\label{fig:entanglement}
\end{figure*}
For two-qubit operations, we consider electric dipole-dipole interaction $V_{dd} = (\mathbf{v}_1 \cdot \mathbf{v}_2R^2 - 3(\mathbf{v}_1 \cdot \mathbf{R})(\mathbf{v}_2 \cdot \mathbf{R}))/4\pi\epsilon R^5$, where $\mathbf{v}_i$ is the spin-dependent charge dipole of qubit $i$ (refs \onlinecite{FlindtPRL2006, GolovachPRB2006}). We note that the orientation of this spin-dependent charge dipole is affected by the top gate since, as discussed before, the effective spin-polarization is modified by the vertical electric field, see Fig. \ref{fig:sketch}(b). For two acceptor qubits $a$ and $b$ separated in the acceptor $z$-plane by a distance $R$ and relative orientation $(\cos\theta_E,\sin\theta_E)$, this electric dipole-dipole interaction works in the two-qubit subspace as $H_{dd}=J_{dd}(\sigma^a_+ +\sigma^a_-)(\sigma^b_+ + \sigma^b_-)$. The dipole-dipole coupling $J_{dd}$ is inversely proportional to the qubit-upper branch energy separation $\Delta$, and it is directly proportional to the qubit-upper branch couplings $\varepsilon_{Zo}$ of each qubit, and their Rashba couplings $\alpha$, such that
\begin{equation}
J_{dd}=\frac{\alpha^a\alpha^b\varepsilon_{Zo}^a\varepsilon_{Zo}^b}{8\pi\epsilon R^3 \Delta ^a \Delta ^b}G(F_z^a,F_z^b,\phi,\theta_E),
\label{eq:J}
\end{equation}
with $G(F_z^a,F_z^b,\phi,\theta_E)$ a modulating function, related to the spin-dependent charge dipoles, that depends on the operating point of each qubit, the magnetic field orientation $\phi$, and $\theta_E$. As discussed previously \cite{KloeffelPRB2013}, adiabatically tuning the gate to turn on/off an electric dipole $\mathbf{v}_i$ allows to modulate the two-qubit coupling\cite{SalfiNano2016}, but this requires the deactiviation of one of the qubits. Introducing the new degree of freedom $\phi$, it is possible to have $J_{dd}$ vanish for particular parameter choices, independently of the modulus of $\mathbf{v}_i$.

Since $J_{dd}$ decays with $\varepsilon_{Zo}^2$, same as $1/T_1$, the number of two-qubit operations per qubit lifetime is constant for any $\phi$, provided that phonons limit the qubit coherence as expected at the sweet spot. Hence, the coupling $J_{dd}$ can be suppressed by choosing parameters such that $G(F_z^a,F_z^b,\phi,\theta_E)=0$ (see purple areas in Fig.~\ref{fig:entanglement}), without deactivating single qubit operations. This condition is the cornerstone of our protocols and generalizes the concept of magic angles\citep{SousaPRA2004} where dipole-dipole interactions vanish.  A conventional magic angle is the angle subtending the magnetic field and relative qubit position, when magnetic dipole-dipole interactions vanish.  Its main drawback is that magnetic field cannot be rapidly swept.  In our case, the magic angle describes the direction of the qubit's spin-dependent electric dipole relative to the qubit positions where spin-dependent electric dipole-dipole interactions vanish. Our protocols are defined at {\em fixed} values of $\phi$. We consider three different situations: (i) both qubits at their isotropic sweet spots; (ii) one qubit is at the isotropic sweet spot and the other at the anisotropic one; (iii) both qubits at their anisotropic sweet spots. Note that gate voltages alone can switch between these cases.

In case (i) the dipolar coupling is perfectly isotropic $G(F_z^a,F_z^b,\phi,\theta_E)=1$, hence neighboring qubits at isotropic sweet spots will always be coupled. In case (ii), see Fig.~\ref{fig:entanglement}(a), we distinguish two interesting orientations in the first quadrant: $\phi=15^{\circ}$ and $75^{\circ}$. These two cases have opposite behavior in the $x$ and $y$ directions: with $\phi=15^{\circ}$ the couplings in the $x$ ($y$)  direction is minimized (maximized), while  $\phi=75^{\circ}$ works in the opposite way. In case (iii), see Fig.~\ref{fig:entanglement}(b), a similar behavior appears for $\phi=12^{\circ}$, $78^{\circ}$ and also for $\phi=40^{\circ}$ and $50^{\circ}$.

We propose two possible protocols for fully tuning two-qubit coupling \textit{electrically} in a rectangular array of acceptors arranged in the $x$ and $y$ directions of the Si lattice, see Figures \ref{fig:entanglement}(c) and (d). Due to the opposite behavior of the coupling in Eq.~\ref{eq:J} for these directions, both protocols require a much larger spacing between acceptors in the direction where $J_{dd}$ is not suppressed, where qubit coupling would be performed by cQED~\citep{KloeffelPRB2013}, while faster dipolar interactions are performed in the perpendicular direction, where they can be switched on and off. Whenever dipolar coupling is switched off, single qubit operations are performed. This general idea can be achieved by fixing the in-plane magnetic field orientation to a particular magic angle. \textbf{Protocol 1}: choose $\phi=12^{\circ}$ or $\phi=40^{\circ}$ ($\phi=78^{\circ}$ or $\phi=50^{\circ}$) and locate acceptors in a close range ($\approx 20$ nm apart for $10^4$ two-qubit operations per qubit lifetime) in the $x$ ($y$) direction and a longer separation in the $y$ ($x$) direction. Sweeping the local gates such that two neighboring qubits in the $x$ ($y$) direction are taken to their anisotropic sweet spots (case (iii)), the coupling $J_{dd}$ is suppressed in that particular direction and each qubit can be addressed individually. Then, taking adiabatically both qubits to the isotropic sweet spot, the coupling is reactivated. In the perpendicular direction, $J_{dd}$ cannot be turned off at this particular $\phi$ so acceptors need to be more separated and entanglement between any pair of qubits is performed via cQED.~\citep{SalfiPRL2016} \textbf{Protocol 2}: in a similar 2D array, choose $\phi=15^{\circ}$  ($\phi=75^{\circ}$).  Sweeping the local gates such that two neighboring qubits in the $x$ ($y$) direction are taken to the isotropic-anisotropic sweet spot combination (case (ii)), the coupling $J_{dd}$ is suppressed in that particular direction and each qubit can be addressed individually. The qubit in the anisotropic sweet spot is adiabatically swept to the isotropic sweet spot, reactivating $J_{dd}$. cQED is performed in the $y$ ($x$) direction.

Both protocols require adiabatically sweeping the vertical electric field to move between the two sweet spots. During the adiabatic sweep the qubit frequency changes making qubits momentarily susceptible to charge noise. The potential for decoherence allows to differentiate the two protocols.  Angles $\phi=40^\circ$ and $\phi=50^\circ$ in Protocol 1 imply a magnetic field orientation very close to the DFS of the anisotropic sweet spot, which means single-qubit operations per qubit lifetime are extremely enhanced. However, the exposure to charge noise in the adiabatic sweeping is higher than in protocol 2 due to the difference in Larmor frequency between sweet spots. Angles $\phi=12^\circ$ and $\phi=78^\circ$ in Protocol 1 and the angles used in Protocol 2 are not close to the DFS, hence the single-qubit operations per time are not particularly enhanced, though $T_1$ is still enhanced with respect to the $\phi=0$ case. Exposure to decoherence during the adiabatic sweep is strongly minimized since the value of $\phi$ is close to the one that merges the isotropic and anisotropic sweet spots (they are barely separated by $\approx$ 1.5 MV/m).

Conventionally, there are two categories of approaches to turn on and off a two-qubit gate: One type achieves tuning of qubit interaction by the reduction of single-qubit dipoles \citep{GolovachPRB2006, SalfiPRL2016}, which means single-qubit operations would also slow down.  Alternatively, two-qubit gates can be effectively halted by detuning the frequencies of the two qubits \citep{TrifPRB2008, KloeffelPRB2013}, however in this approach two-qubit dynamics is only suppressed, not turned off.  Since in both our protocols two-qubit couplings can be totally suppressed, our proposal here is clearly a more efficient alternative in precisely controlling two-qubit operations.

We also note that both protocols are robust against acceptor placement imprecision. As an example, for an accuracy of  $\pm 5$ nm in-plane position, and inter-acceptor distance of $20$ nm, the angle variation would be of $\pm 15^{\circ}$. Under these circumstances, $J_{dd}$, although non-zero, would be significantly suppressed (see Fig.~\ref{fig:entanglement}).

 In summary, we have found an unexpected magnetic field orientation dependence of all the parameters involved in one and two-qubit operations with acceptors in Si. This dependence can be used to develop tailored protocols at fixed in-plane magnetic field orientations (magic angles) that allow to perform one and two-qubit operations independently, by purely electrical means, while maintaining each qubit in sweet spots where charge noise effects are suppressed. The proposed protocols are well within reach using state-of-the-art technology, paving the way for a full electrically controlled Si-based quantum computer implementation.

\vspace{0.2cm}\noindent\textbf{ACKNOWLEDGMENTS}

\noindent JCAU and MJC acknowledge funding from Ministerio de Econom\'ia, Industria y Competitividad (Spain) via Grants No FIS2012-33521 and FIS2015-64654-P and from CSIC (Spain) via grant No 201660I031. JCAU thanks the support from grants BES-2013-065888 and EEBB-I-16-11046. DC and SR are supported by the Australian Research Council Centres of Excellence FLEET (CE170100039) and CQC2T (CE110001027), respectively.  XH thanks support by US ARO through grant W911NF1210609, and Gordon Godfrey Fellowship from UNSW School of Physics. JS acknowledges financial support from an ARC DECRA fellowship (DE160101490).

\bibliography{acceptors}

\begin{thebibliography}{62}
\bibitem[{not({\natexlab{a}})}]{note1}
\bibinfo{note}{Such an interaction is not only important for direct control of
  the qubit, but also essential if coupling to a superconductor resonator is
  sought~\citep{PalyiPRL2012, KloeffelPRB2013, TrifPRB2008, XuedongPRB2012,
  ChildressPRA2004}}.
\end{thebibliography}
\newpage
\beginsupplement
\begin{widetext}
\centering\noindent\textbf{SUPPLEMENTARY INFORMATION}

\section{Kohn-Luttinger and Bir-Pikus Hamiltonians}
The Kohn-Luttinger Hamiltonian for the valence band of semiconductors \citep{KohnPR1955}, including the Coulomb impurity, is 
\begin{equation}H_{\rm KL}=
\begin{pmatrix}
P+Q & L & M & 0 & \frac{i}{\sqrt{2}}L & -i\sqrt{2}M \\
L^*  &  P-Q & 0 & M & -i\sqrt{2}Q & i\sqrt{\frac{3}{2}}L \\
M^* & 0 & P-Q & -L & -i\sqrt{\frac{3}{2}}L^* & -i\sqrt{2}Q \\
0  & M^* & -L^* & P+Q & -i\sqrt{2}M^* & -\frac{i}{\sqrt{2}}L^* \\
-i\sqrt{2}L^* & i\sqrt{2}Q & i\sqrt{\frac{3}{2}}L & i\sqrt{2}M & P+\Delta_{SO} & 0 \\
i\sqrt{2}M^*  & -i\sqrt{\frac{3}{2}}L^* & i\sqrt{2}Q & i\sqrt{2}L & 0 & P+\Delta_{SO}
\end{pmatrix} 
\label{HKL}
\end{equation}
We can define the effective Rydberg unit as $Ry^*=e^4m_0/2\hbar^2\epsilon_s^2\gamma_1$ and the effective Bohr radius as $a^*=\hbar^2\epsilon_s\gamma_1/e^2m_0$. In these units the differential operators in Eq.~(\ref{HKL}) are 
\begin{eqnarray}
P&=&-k^2+\frac{2}{r} \nonumber \\
Q&=&-\frac{\gamma_2}{\gamma_1}(k_x^2+k_y^2-2k_z^2) \nonumber \\
L&=&i2\sqrt{3}\frac{\gamma_3}{\gamma_1}(k_x-ik_y)k_z   \\
M&=&-\sqrt{3}\frac{\gamma_2}{\gamma_1}(k_x^2-k_y^2)+i2\sqrt{3}\frac{\gamma_3}{\gamma_1}k_x k_y  \nonumber \, ,
\label{PQLM}
\end{eqnarray}
with $m_0$ the free electron mass, $\epsilon_s$ the semiconductor static dielectric constant, and  $\gamma_1$, $\gamma_2$ and $\gamma_3$ material dependent Luttinger parameters.
The interaction with strain is given by the Bir-Pikus Hamiltonian \citep{BirJPCS1963}:
\begin{eqnarray}
H_{\rm BP}&=&a\epsilon\mathbb{1}\nonumber \\ &+&b\left((J_x^2-\frac{5}{4}\mathbb{1})\epsilon_{xx}+(J_y^2-\frac{5}{4}\mathbb{1})\epsilon_{yy} 
+ (J_z^2-\frac{5}{4}\mathbb{1})\epsilon_{zz}\right) \nonumber \\
&+&d/\sqrt{3}\left(\{J_x,J_y\}\epsilon_{xy}
+\{J_y,J_z\}\epsilon_{yz}+\{J_x,J_z\}\epsilon_{xz}\right)
\label{HBP}
\end{eqnarray}
The parameters $a,b$ and $d$ are the deformation potentials of the host material, and $\epsilon_{ij}$ are the deformation tensor components.

\section{Effective $4\times 4$ Hamiltonian}
The results of the diagonalization of the total Hamiltonian are mapped onto an effective Hamiltonian for the four lowest states. For any magnetic field orientation $\phi$ the Hamiltonian in the $|m_J\rangle$ basis is \citep{SalfiPRL2016}:
\begin{equation}
H_{\rm eff}=\begin{pmatrix}
0 & \frac{\sqrt{3}}{2}\varepsilon_Ze^{-i\phi} & -ipF_z & 0 \\
\frac{\sqrt{3}}{2}\varepsilon_Ze^{i\phi}  & \Delta_{HL} & \varepsilon_Ze^{-i\phi} & -ipF_z \\
ipF_z & \varepsilon_Ze^{i\phi} & \Delta_{HL} & \frac{\sqrt{3}}{2}\varepsilon_Ze^{-i\phi} \\
0 & ipF_z  & \frac{\sqrt{3}}{2}\varepsilon_Ze^{i\phi} & 0
\end{pmatrix} \,.
\end{equation}
For the qubit Hamiltonian, we define $E_l=\frac{1}{2}(\Delta_{HL}-\sqrt{\Delta_{HL}^2+4p^2F_z^2})$, $E_u=\frac{1}{2}(\Delta_{HL}+\sqrt{\Delta_{HL}^2+4p^2F_z^2})$, $a_L=E_l/\sqrt{E_l^2+p^2F_z^2}$ and $a_H=pF_z/\sqrt{E_l^2+p^2F_z^2}$. After the transformations shown in Ref. \citep{SalfiPRL2016} the qubit Hamiltonian is: 
\begin{equation}
H_{\rm qubit}=\begin{pmatrix}
E_l-\frac{1}{2}\varepsilon_{Zl} & 0 & Z_1 & Z_2 \\
0 & E_l+\frac{1}{2}\varepsilon_{Zl} & Z_2 & Z_1 \\
Z_1 & -Z_2 & E_u-\frac{1}{2}\varepsilon_{Zu} & 0 \\
-Z_2 & Z_1 & 0 & E_u-\frac{1}{2}\varepsilon_{Zu}
\end{pmatrix}
\end{equation}
being 
\begin{eqnarray}
Z_1&=&\frac{1}{2}\varepsilon_{Zo}\cos(\theta_l/2-\theta_u/2-\theta_o) 	\nonumber  \\
Z_2&=&\frac{i}{2}\varepsilon_{Zo}\sin(\theta_l/2-\theta_u/2-\theta_o).
\end{eqnarray}
Where the first two states correspond to the qubit branch and the last two to the upper branch. The different relevant couplings and the associated phases are defined as follows
\begin{eqnarray}
\theta_l&=&\arctan(a_L^2\cos(\phi)+\sqrt{3}a_La_H\sin(\phi),-a_L^2\sin(\phi)-\sqrt{3}a_La_H\cos(\phi)) \nonumber
\\
\theta_u&=&\arctan(-a_H^2\cos(\phi)+\sqrt{3}a_La_H\sin(\phi),-a_H^2\sin(\phi)+\sqrt{3}a_La_H\cos(\phi))
\\
\theta_o&=&\arctan(-a_La_H\sin(\phi)+\sqrt{3}/2(a_L^2-a_H^2)\cos(\phi),-a_La_H\cos(\phi)+\sqrt{3}/2(a_L^2-a_H^2)\sin(\phi)) \nonumber
\\
\varepsilon_{Zl}&=&2\varepsilon_Z\sqrt{3a_L^2a_H^2+a_L^4+2\sqrt{3}a_L^3a_H\sin(2\phi)} \nonumber
\\
\varepsilon_{Zu}&=&2\varepsilon_Z\sqrt{3a_L^2a_H^2+a_H^4-2\sqrt{3}a_H^3a_L\sin(2\phi)} 
\\
\varepsilon_{Zo}&=&2\varepsilon_Z\sqrt{3(a_L^2-a_H^2)/4+a_H^2a_L^2+\sqrt{3}a_Ha_L(a_H^2-a_L^2)\sin(2\phi)} 	\nonumber
\end{eqnarray}
Physically, $\varepsilon_{Zl}$ and $\varepsilon_{Zu}$ are the qubit and upper Zeeman splittings while $\varepsilon_{Zo}$ is the qubit-upper branch coupling. The phases associated to these couplings $\theta_{l}$, $\theta_{u}$, $\theta_{o}$ can be related to an effective spin polarization of each Kramer doublet and the interacting term respectively.

\section{Sweet spots}
The value of the qubit Larmor frequency is given, to first order, by $\varepsilon_{Zl}$. Its value depends explicitly on both the electric field and magnetic field magnitudes, but  also depends on the magnetic field orientation. We can look for sweet spots by simply finding the solutions to $d\varepsilon_{Zl}/dF=0$:
\begin{equation}
\frac{d\varepsilon_{Zl}}{dF_z}=\varepsilon_Z\frac{(-3+4a_L^2)(a_H+\sqrt{3}a_L\sin(2\phi))a'_L(F_z)}{a_H\sqrt{3-2a_L^2+2\sqrt{3}a_La_H\sin(2\phi)}}=0 \,.
\end{equation}
One of the solutions corresponds to the fixed sweet spot $a_L=-\sqrt{3}/2$ ($a_H=1/2$). Considering that $a'_L\propto a_H$, the other solution is equivalent to solve $a_H+\sqrt{3}a_L\sin(2\phi)=0$. Considering positive electric fields $a_H\geq 0$ and $a_L\leq 0$, meaning that for $0\leq \phi \leq \pi/2$ and $\pi \leq \phi \leq 3\pi/2$ there is a $\phi$ dependent sweet spot solution.

Particularly, at $\phi=\pi/4+n\pi$ this corresponds to $a_H=\sqrt{3}/2$ and $a_L=-1/2$ while for a magnetic field aligned with the main axes of the crystal, this sweet spot corresponds to the value $F_z=0$.

Also, when $\phi=\pi/4+n\pi$ the value of the Larmor frequency at the isotropic sweet spot is zero. At this point there is an inversion of the effective spin polarization of the lower branch, implying that an effective \textit{g}-factor flip occurs in this particular case.

\section{Decoherence Free Subspace (DFS)}
As the qubit interacts with the upper branch states through the Zeeman interaction, we can see this subspace as a leakage submanifold. The interaction terms between the qubit states and the leakage states are then given by the off-diagonal Zeeman terms in the qubit Hamiltonian. A decoherence and relaxation free subspace would be a subspace in which the Zeeman interaction is purely diagonal in the qubit basis, or equivalently $\left[H_{T_d}+H_{\text{interface}},H_B\right]=0$. In this subspace, all the interactions would be suppressed to first order. To find this subspace we use the following elements that form part of a bigger basis of the space of spin $3/2$:
\begin{eqnarray}
e_1&=&\frac{\mathbb{1}}{2} \nonumber \\
e_2&=&1/6(2J_z^2-J_x^2-J_y^2) \nonumber \\
e_3&=&1/\sqrt{5}J_x \nonumber \\
e_4&=&1/\sqrt{5}J_y \\
e_5&=&1/\sqrt{12}\left\{J_x,J_y\right\} \nonumber \\
e_6&=&1/\sqrt{12}\left\{J_y,J_z\right\} \nonumber \\
e_7&=&1/\sqrt{12}\left\{J_z,J_x\right\} \nonumber
\end{eqnarray}
In this basis the elements of the effective Hamiltonian are
\begin{eqnarray}
H_{\text{inter}}&=&\Delta_{HL}(e_1-e_2) \nonumber \\
H_{T_d}&=&2pF_ze_5 \\
H_B&=&\sqrt{5}(B_xe_3+B_ye_4) \nonumber
\end{eqnarray}
The commutators are then
\begin{eqnarray}
\left[H_{\text{inter}},H_B\right]&=& i2\sqrt{3}\Delta_{HL}(B_ye_7-B_xe_6) \nonumber \\
\left[H_{T_d},H_B\right]&=&-4ipF_z(B_xe_7-B_ye_6)
\end{eqnarray}
The total commutator is
\begin{eqnarray}
\left[H_{\text{inter}}+H_{T_d},H_B\right]&=&e_7(-4ipF_zB_x+i2\sqrt{3}\Delta_{HL}B_y) \nonumber \\
+&e_6&(4ipF_zB_y-i2\sqrt{3}\Delta_{HL}B_x)
\end{eqnarray}
As $e_7$ and $e_6$ are different elements of the basis, $ \left[H_{\text{inter}}+H_{T_d},H_B\right]= 0$ is equivalent to solve the following system
\begin{eqnarray}
\Delta_{HL}B_y-\frac{2pF_z}{\sqrt{3}}B_x=0 	\nonumber \\
\Delta_{HL}B_x-\frac{2pF_z}{\sqrt{3}}B_y=0
\end{eqnarray}
The system has non-trivial solutions if and only if $B_x=\pm B_y$ which corresponds to the orientations $\pm \pi/4+n\pi$. These are the most symmetric directions as the main axes of the crystal remain indistinguishable.

Case $B_x=-B_y$, the solution requires $\Delta_{HL}=-\frac{2pF_z}{\sqrt{3}}$ which corresponds to the isotropic sweet spot. The case $B_x=B_y$ requires $\Delta_{HL}=\frac{2pF_z}{\sqrt{3}}$ corresponding to the anisotropic sweet spot. In these two cases the Zeeman interaction can be diagonalized simultaneously with the interface and $T_d$ symmetry terms, so the qubit becomes isolated from the upper branch.

It is also interesting to express the different contributions in terms of the spherical tensors $J_+$ and $J_-$
\begin{eqnarray}
H_{T_d}+H_{\text{inter}}&=&-\frac{3}{4}\mathbb{1}-\frac{i}{8}(\Delta_{HL}+2pF_z/\sqrt{3})(J_++iJ_-)^2 \nonumber \\ &+&\frac{i}{8}(\Delta_{HL}-2pF_z/\sqrt{3})(J_+-iJ_-)^2 \nonumber \\
H_{B}&=&\frac{1+i}{4}(B_x-B_y)(J_+-iJ_-)  \\ &+&\frac{1-i}{4}(B_x+B_y)(J_++iJ_-) \nonumber
\end{eqnarray}
From here, it can be seen how for the sweet spots, and for particular values of the magnetic fields, the non-magnetic and the magnetic terms share eigenvectors.

\section{Effective Spin polarization induced by the spin-orbit interaction}
The qubit states in the $|m_J\rangle$ basis are:
\begin{equation}
|l\pm\rangle =a_L|\pm 1/2\rangle \mp ia_H|\mp 3/2\rangle
\end{equation}
Transforming the $m_J$ basis into the $|l_z,s_z\rangle$ basis using the Clebsch-Gordan coefficients
\begin{equation}
|l\pm\rangle =\frac{a_L}{\sqrt{3}}(|\pm 1, \mp 1/2\rangle+\sqrt{2}|0, \pm 1/2\rangle) \mp ia_H|\mp 1, \mp 1/2\rangle
\end{equation}
To get the effective spin polarization of this branch we can compute the matrix of expected values for the in-plane spin operators $\langle l'|\sigma_i|l\rangle$:
\begin{equation}
\langle l'|\sigma_x|l\rangle= \frac{a_L}{3}\begin{pmatrix}
0 & a_L+i\sqrt{3}a_H \\
a_L-i\sqrt{3}a_H & 0
\end{pmatrix}
\end{equation}
\begin{equation}
\langle l'|\sigma_y|l\rangle= \frac{a_L}{3}\begin{pmatrix}
0 & ia_L+\sqrt{3}a_H \\
-ia_L+\sqrt{3}a_H & 0
\end{pmatrix}
\end{equation}
At the isotropic sweet spot $a_L=-\sqrt{3}/2$ and $a_H=1/2$, making $\langle l'|\sigma_x|l\rangle=-\langle l'|\sigma_y|l\rangle$, and at the same time the expected value at the direction $\phi=-\pi/4$ becomes saturated to the maximum possible value of spin projection  $1/2$ which is equivalent to an effective spin polarization in the $\phi=-\pi/4\pm \pi$ direction. The change of effective spin-polarization in the qubit branch by rotating the magnetic field can be seen in the supplemental file $gs\_spin\_ polarization.mov$.
The other interesting possibility is $a_L=-1/2$ and $a_H=\sqrt{3}/2$ which corresponds to the anisotropic sweet spot in the particular case $\phi=\pi/4$. In this case the upper states have the composition $a_L=-\sqrt{3}/2$  and $a_H=1/2$ so $\langle u'|\sigma_x|u\rangle=\langle u'|\sigma_y|u\rangle$ indicating an effective spin polarization of the upper branch for $\phi=\pi/4\pm \pi$.

\section{Phonon-induced Spin Relaxation}
At low temperatures the expression for the relaxation times of the acceptor qubit is
\begin{equation}
\frac{1}{T_1}=\frac{(\hbar \omega)^3}{20\hbar^4 \pi \rho}\Big[ \sum_i |\langle -|D_{ii}|+\rangle |^2(\frac{2}{v_l^5}+\frac{4}{3v_t^5})+ \sum_{i\neq j} |\langle -|D_{ij}|+ \rangle |^2(\frac{2}{3v_l^5}+\frac{1}{v_t^5})\Big]
\end{equation}
Going to second order in the Schrieffer-Wolff transformation~\citep{Winklerbook} we get
\begin{equation}
\langle -|D_{ij}|+ \rangle = \frac{1}{E_l-E_u}(\hat{H}'_{-,u-}\hat{H}'_{u-,+}+\hat{H}'_{-,u+}\hat{H}'_{u+,+})
\end{equation}
Where $\hat{H}'=\hat{H}'_{Z}+\hat{H}'_{ph}$. The elements of the Hamiltonian $\hat{H}'_{ph}$ are
\begin{eqnarray}
D_{ii}&=&b'(J_i^2-\frac{5}{4}) \nonumber \\
D_{ij}&=&2d'/\sqrt{3}\{J_i,J_j\} \ \ i\neq j
\end{eqnarray}
The values of $|\langle -|D_{ij}|+ \rangle|^2$ at the isotropic sweet spot are independent of the angle $\phi$ except for the intrinsic dependence of $\varepsilon_{Zo}$. These values are 
\begin{eqnarray}
|\langle -|D_{xx}|+ \rangle|^2&=&|\langle -|D_{yy}|+ \rangle|^2=3b^2\varepsilon_{Zo}^2/64\Delta^2 \nonumber \\ 
|\langle -|D_{zz}|+ \rangle|^2&=&3b^2\varepsilon_{Zo}^2/16\Delta^2 \nonumber \\
|\langle -|D_{xy}|+ \rangle|^2&=&d^2\varepsilon_{Zo}^2/16\Delta^2 \\
|\langle -|D_{xz}|+ \rangle|^2&=&|\langle -|D_{yz}|+ \rangle|^2=d^2\varepsilon_{Zo}^2/8\Delta^2 \nonumber 
\end{eqnarray}

In the case of the anisotropic sweet spot the values of $|\langle -|D_{ij}|+ \rangle|^2$ are 
\begin{eqnarray}
|\langle -|D_{xx}|+ \rangle|^2&=&|\langle -|D_{yy}|+ \rangle|^2=3b^2\varepsilon_{Zo}^2/64\Delta^2 \nonumber \\ 
|\langle -|D_{zz}|+ \rangle|^2&=&3b^2\varepsilon_{Zo}^2/16\Delta^2 \nonumber \\ 
|\langle -|D_{xy}|+ \rangle|^2&=&d^2\varepsilon_{Zo}^2/16\Delta^2 \\
|\langle -|D_{xz}|+ \rangle|^2&=&d^2\varepsilon_{Zo}^2\cos^2\theta_o/4\Delta^2 \nonumber \\ 
|\langle -|D_{yz}|+ \rangle|^2&=&d^2\varepsilon_{Zo}^2\sin^2\theta_o/4\Delta^2 \nonumber
\end{eqnarray}
Collecting terms, the new dependence on the phase $\theta_o$ cancels out so we arrive to the same formula except for different definitions of $\varepsilon_{Zo}$ and the energy difference $\Delta=E_l-E_u$.

The general formula for phonon-induced relaxation at the sweet spots is then
\begin{equation}
\frac{1}{T_1}=\frac{(\hbar \omega(\phi))^3}{20\hbar^4 \pi \rho}\left(\frac{\varepsilon_{Zo}(\phi)}{E_l-E_u}\right)^2 \Big[ \frac{3b'}{32}(\frac{2}{v_l^5}+\frac{4}{3v_t^5})+\frac{5d'}{48}(\frac{2}{3v_l^5}+\frac{1}{v_t^5})\Big]
\end{equation}

\section{Single qubit operations}
The lack of inversion symmetry allows the action of an in-plane linear Stark effect $H_E=e(E_xx+E_yy)=eE_\parallel(\cos\theta_\parallel+\sin\theta_\parallel)$, see Ref.~\citep{SalfiPRL2016}. The effect of this interaction of the acceptor hole is calculated considering several excited states and mapped onto the effective $4\times 4$ Hamiltonian. In the qubit basis it becomes
\begin{equation}
\hat{H}_{E}=\begin{pmatrix}
0 & 0 & E_{1R} & E_{2R} \\
0 & 0 & E_{2R} & E_{1R} \\
-E_{1R} & E_{2R} & 0 & 0 \\
E_{2R} & -E_{1R} & 0 & 0
\end{pmatrix}
\end{equation}
being
\begin{eqnarray}
E_{1R}&=&i\alpha E_{\parallel}\sin(\theta_{\parallel}+\theta) \nonumber \\
E_{2R}&=&-\alpha E_{\parallel}\cos(\theta_{\parallel}+\theta)
\end{eqnarray}
Where $\theta=\theta_u/2-\theta_l/2$.
Applying a second order SW transformation~\citep{Winklerbook, SalfiPRL2016} we get that the qubit EDSR term is
\begin{equation}
D=\alpha\frac{\varepsilon_{Zo}(\phi)}{E_l-E_u}\cos(\theta_o-\theta_\parallel) \,.
\end{equation}
With this coupling and the limiting factor $T_1$ we can calculate the number of $\pi$ rotations per qubit lifetime, see Fig.~\ref{fig:ops}. The number of qubit operations diverges when $T_1$ tends to infinity, since $T_1\propto 1/\varepsilon_{Zo}^2$ while $D\propto \varepsilon_{Zo}$.
\begin{figure}
\includegraphics[clip,width=0.5\textwidth]{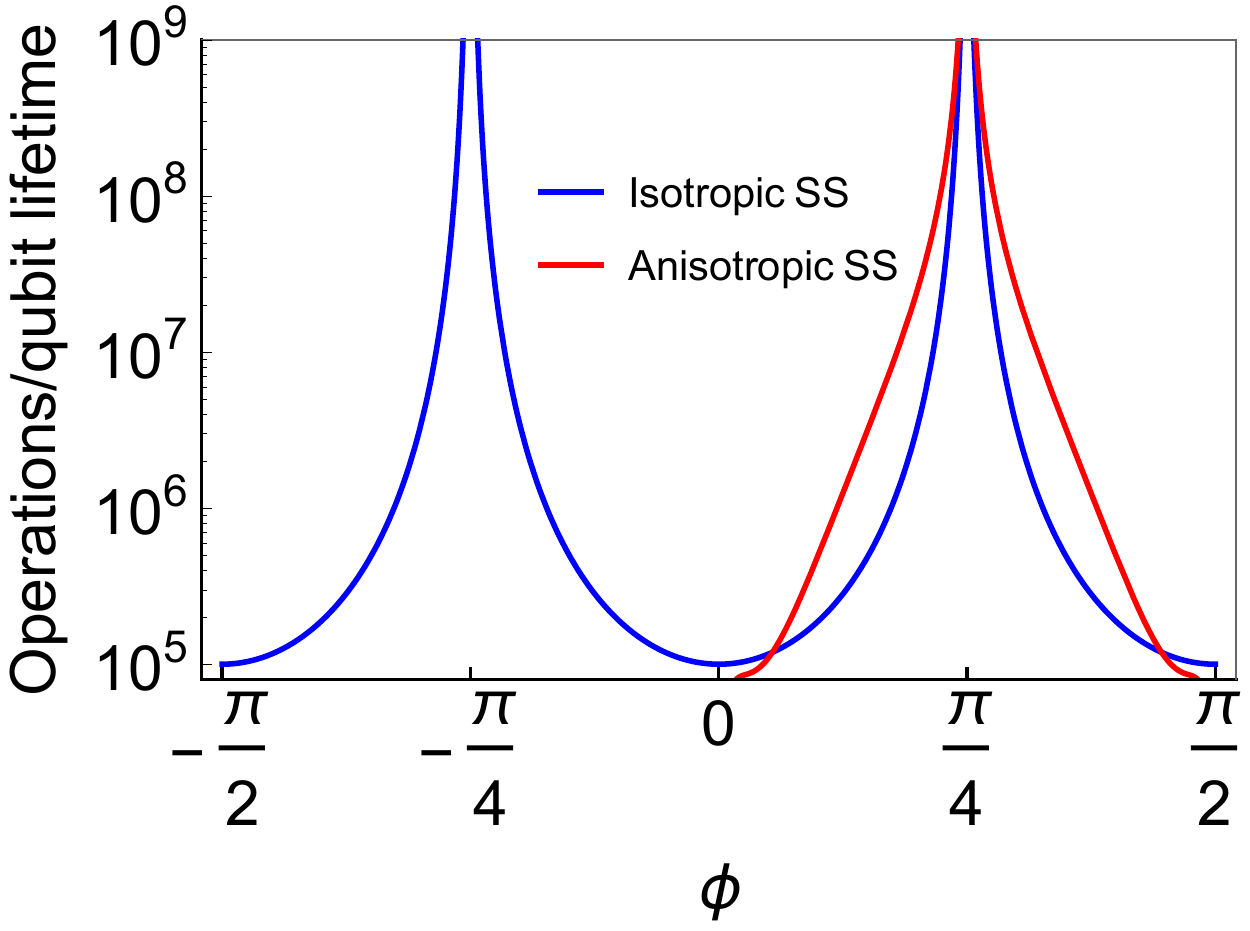}
\caption{Number of single qubit operations per qubit lifetime as a function of $\phi$ for an acceptor at 4.6 nm from the interface and $B=0.5$T.}
\label{fig:ops}
\end{figure}
\section{Two qubit coupling}
Consider two acceptors separated by a distance $\mathbf{R}$. Due to the spin-orbit interaction and the electric dipole moment of each acceptor, a spin-dependent dipolar interaction is expected between the two acceptors. Let the subspace of the two qubits be $\{|l-^a,l-^b\rangle, |l-^a,l+^b\rangle, |l+^a,l-^b\rangle, |l+^a,l+^b\rangle\}$. The total Hamiltonian is $H^\Sigma=H_{op}^a+H_{op}^2+V^{12}$, where $H_{op}$ are the single acceptor Hamiltonians and $V^{12}$ is the Hamiltonian of the electrostatic interaction given by $V^{12}(\mathbf{r}_1-\mathbf{r}_2)=e^2/4\pi\epsilon|\mathbf{r}_1-\mathbf{r}_2|$.

Here we assume that each qubit may have different energies and applied fields. The single qubit Hamiltonians are
\begin{equation}
\langle m|H_{op}^i|m'\rangle = \left(
\begin{array}{cccc}
 -\frac{\varepsilon_l^i}{2} & 0 & Z_1^i & i Z_2^i \\
 0 & \frac{\varepsilon_l^i}{2} & i Z_2^i & Z_1^i \\
 Z_1^i & -i Z_2^i & \Delta ^i-\frac{\varepsilon_u^i}{2} & 0 \\
 -i Z_2^i & Z_1^i & 0 & \frac{\varepsilon_u^i}{2}+\Delta ^i \\
\end{array}
\right)
\end{equation}
where the superindex $i$ indicates acceptor a or b.

When the two acceptors are far enough we can use the multi-pole expansion for the Coulomb interaction:
\begin{equation}
\langle mn|V^{12}|m'n'\rangle = \dfrac{R^2\langle m|e\mathbf{r'_1}|m'\rangle \cdot \langle n|e\mathbf{r'_2}|n'\rangle-3(\langle m|e\mathbf{r'_1}|m'\rangle \cdot \mathbf{R})(\langle n|e\mathbf{r'_2}|n'\rangle \cdot \mathbf{R})}{4\pi\epsilon R^5}
\end{equation}
Being $\mathbf{r'}_i=\mathbf{r}_i-\mathbf{R}_i$ the hole coordinate relative to the ion, and assuming an arbitrary relative position in the $xy$ plane $\mathbf{R}=R\cos(\theta_E)\hat{x}+R\sin(\theta_E)\hat{y}$, so
\begin{equation}
\langle mn|V^{12}|m'n'\rangle = \dfrac{(1-3\cos^2\theta_E)\langle m|e x'_1|m'\rangle \cdot \langle n|e x'_2|n'\rangle+(1-3\sin^2\theta_E)\langle m|e y'_1|m'\rangle \cdot \langle n|e y'_2|n'\rangle+\langle m|e z'_1|m'\rangle \cdot \langle n|e z'_2|n'\rangle}{4\pi\epsilon R^3}
\end{equation}
The dipole matrix elements relevant for the Coulomb interaction are
\begin{equation}
\langle m|e(x',y')^i|m'\rangle = \left(
\begin{array}{cccc}
 0 & 0 & i q_{1x,y}^i & q_{2x,y}^i \\
 0 & 0 & q_{2x,y}^i & iq_{1x,y}^i \\
 -iq_{1x,y}^i & q_{2x,y}^i & 0 & 0 \\
 q_{2x,y}^i & -iq_{1x,y}^i & 0 & 0 \\
\end{array}
\right)
\end{equation}
where $q_{1x}^i=\alpha^i\sin\theta^i$, $q_{2x}^i=-\alpha^i\cos\theta^i$, $q_{1y}^i=\alpha^i\cos\theta^i$, and $q_{2y}^i=\alpha^i\sin\theta^i$ (assuming the approximation $\alpha\gg p$ valid for both qubits).

We can project the interactions into the $4\times 4$ subspace using a SW transformation. Working out the second order correction $H^{(2)}$ we get a spin-independent shift
\begin{equation}
H^{(2)}=-I\left[\frac{\left(\left(q_1^a\right)^2+\left(q_2^a\right)^2\right)
   \left(\left(q_1^b\right)^2+\left(q_2^b\right)^2\right)}{\Delta ^a+\Delta
   ^b}+\frac{\left(Z_1^a\right)^2+\left(Z_2^b\right)^2}{\Delta^a}+\frac{\left(Z_1^b\right)^2+\left(Z_2^b\right)^2}{\Delta ^b}\right]
\end{equation}
The third-order correction is 
\begin{eqnarray}
H^{(3)}_{mm'}=-\frac{1}{2}\sum_{l,m''}\left[\frac{H'_{ml}H'_{lm''}H'_{m''m}}{(E_{m'}-E_l)(E_{m''}-E_l)}+\frac{H'_{mm''}H'_{m''l}H'_{lm'}}{(E_{m}-E_l)(E_{m''}-E_l)}\right] \nonumber \\
+\frac{1}{2}\sum_{l,l'}H'_{ml'}H'_{ll'}H'_{l'm'}\left[\frac{1}{(E_m-E_l)(E_m-E_{l'})}+\frac{1}{(E_{m'}-E_l)(E_{m'}-E_{l})}\right] \,.
\end{eqnarray}
The result of the third order correction to zeroth order in $\varepsilon_i$ is
\begin{equation}
H^{(3)}=J_{dd}\begin{pmatrix}
0 & 0 & 0 & 1 \\
0 & 0 & 1 & 0 \\
0 & 1 & 0 & 0 \\
1 & 0 & 0 & 0
\end{pmatrix}
\end{equation}
where
\begin{equation}
J_{dd}=\frac{4 \left(q_2^a Z_1^a+q_1^a Z_2^a\right)
   \left(q_2^b Z_1^b+q_1^b Z_2^b\right)}{\Delta ^a \Delta ^b}
\end{equation}
which is an Ising type spin-spin interaction $H^{(3)}=J_{dd}(\sigma_{a+}+\sigma_{a-})(\sigma_{b+}+\sigma_{b-})$.

After adding all the contributions and substituting, the value of $J_{dd}$ is:
\begin{equation}
J_{dd}=\frac{\alpha^a \alpha^b \varepsilon_{Zo}^a \varepsilon_{Zo}^b (\sin \theta_{o}^a \sin \theta_{o}^b (1-3 \sin^2 \theta_{E})+\cos \theta_{o}^a \cos  \theta_{o}^b (1-3 \cos^2 \theta_{E}))}{4 \pi  \epsilon  R^3 (E_l^a-E_u^a) (E_l^b-E_u^b)}
\end{equation}
This expression can be substituted for each case

\textbf{Case (i)}:
We get
\begin{equation}
J_{dd}=\frac{3}{2}\varepsilon_Z^2\left(1+\sin(2\phi)\right)\frac{\alpha^a \alpha^b}{8 \pi  \epsilon  R^3 (E_l^a-E_u^a) \,.
   (E_l^2-E_u^2)}
\end{equation}

\textbf{Case (ii)}:
Omitting the complex dependence in the dynamic sweet spot of $\theta_o$ on $\phi$
\begin{equation}
J_{dd}=3\varepsilon_Z^2\sqrt{1+\sin(2\phi)}\frac{\alpha^a \alpha^b |\cos(2\phi)|^2(3\cos(3\phi-\theta_o^a)-\cos(\phi+\theta_o^a)-3\cos(2\theta_E)(\cos(\phi-\theta_o^a)-3\cos(3\phi+\theta_o^a))\big]}{16 \pi  \epsilon  R^3 (E_l^a-E_u^a)
   (E_l^b-E_u^b)\sqrt{7+4\cos 4\phi-3\cos 8\phi}} \,.
\end{equation}

\textbf{Case (iii)}:
\begin{equation}
J_{dd}=3\varepsilon_Z^2\frac{\alpha^a \alpha^b \cos^2(2\phi)\big[(1-3\cos^2\theta_E)\cos\theta_o^a\cos\theta_o^b+(1-3\sin^2\theta_E)\big]\sin\theta_o^a\sin\theta_o^b}{8 \pi  \epsilon  R^3 (E_l^a-E_u^a)
   (E_l^b-E_u^b)(3\cos(4\phi)-5)} \,.
\end{equation}
The normalized angular distribution of cases (ii) and (iii) as a function of the relative orientation $\theta_E$, and the magnetic field orientation can be seen in the supplementary movie $two-qubit-coupling-distribution.mov$. Where the blue and red curves correspond to case (ii) and (iii) respectively.
\section{Charge noise exposure during the entanglement protocols}
Since the exact amount of charge noise is device dependent, we account for the charge noise exposure qualitatively. For $\phi=n\pi/2$ as in~\citep{SalfiPRL2016, SalfiNano2016}, the electric field is adiabatically swept from $F_z=0$ to $F_z$ at the isotropic sweet spot, hence the simplest qualitative way of comparing protocols is to account for the ratio $T_2^*(\phi)/T_2^*(0)$. Here $T_2^*(\phi)$ would be related to the amount of charge noise accumulated when going from the anisotropic sweet spot to the isotropic sweet spot for a given $\phi$, and $T_2^*(0)$ the exposure to charge noise for $\phi=0$.
 
\begin{figure}[t!]
\includegraphics[clip,width=0.5\textwidth]{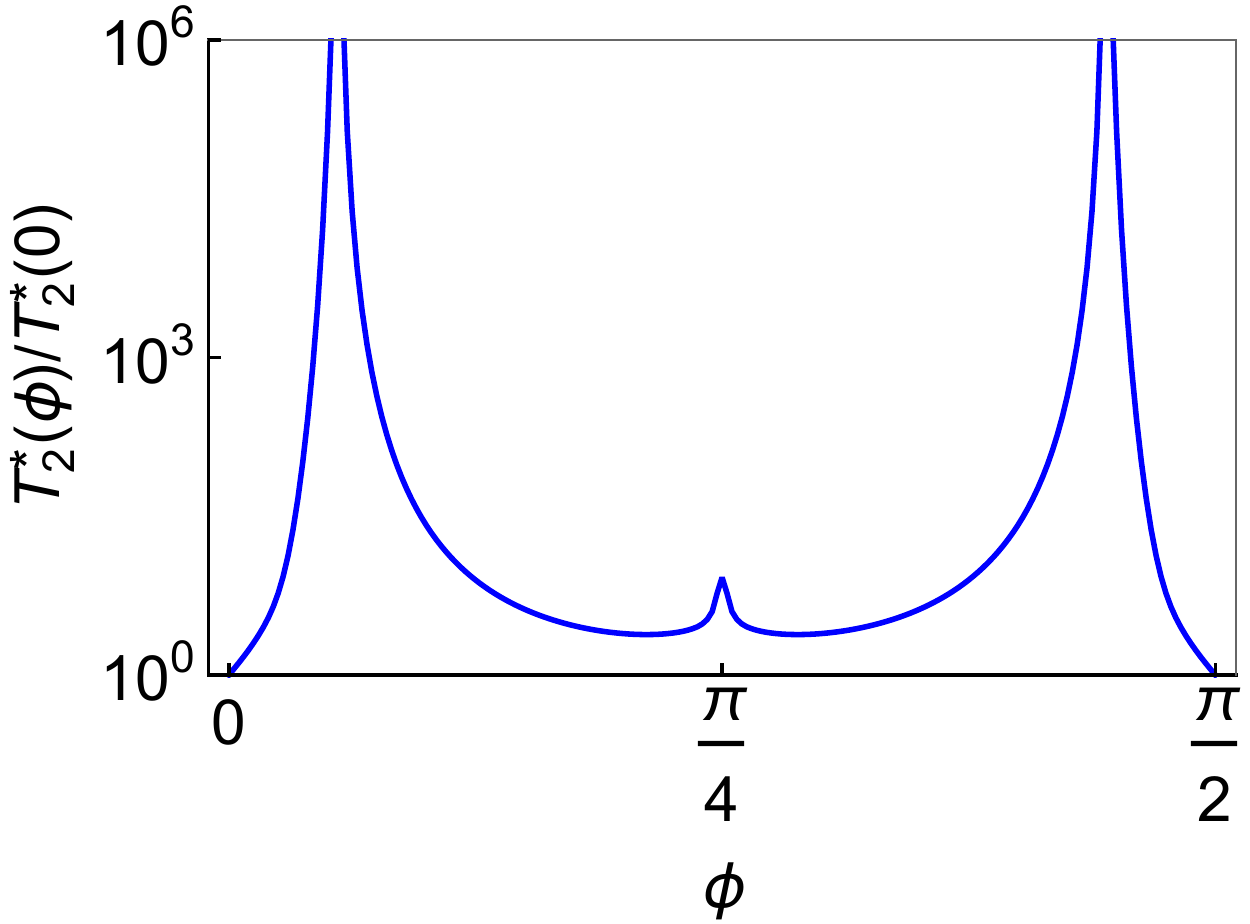}
\caption{Ratio $T_2^*(\phi)/T_2^*(0)$ as a function of $\phi$. Note that the charge noise exposure is always better for any $\phi\neq 0$ as $T_2^*(\phi)/T_2^*(0)\geq 1$.}
\label{fig:T2path}
\end{figure}

Given a fluctuating charged defect with field $\delta F_z$ we compute the change in energy $\delta E_{\rm Larmor}$ for a given value of $F_z$ in the qubit. How much the defect affects the qubit energy depends on the derivative of the Larmor energy on $F_z$. From \citep{DimiAPL2009} we know that $1/T_2^*\propto \delta E_{\rm Larmor}^2$. Defining $F_z^*$ and $\tilde{F}_z$ as the values of the vertical electric field at the isotropic and anisotropic sweet spots respectively, we account for the total charge noise exposure by integrating the Larmor energy change along the path from $\tilde{F}_z$ to $F_z^*$
\begin{equation}
I(\phi)=\int_{\tilde{F}_z}^{F_z^*}\delta E_{\rm Larmor}^2(F_z) dF_z
\end{equation}
By assuming a constant sweep rate, the time of exposure to charge noise is proportional to the difference between the initial and final electric fields. In total we get
\begin{equation}
\frac{T_2^*(\phi)}{T_2^*(0)}=\mid\frac{I(0)F_z^*}{I(\phi)(F_z^*-\tilde{F}_z)}\mid
\end{equation}
Intuitively, this ratio is simply proportional to the charge noise sensitivity along the path and its length.  The values of this ratio can be seen in Fig.~\ref{fig:T2path}.

From Fig.~\ref{fig:T2path} it is clear that any $\phi\neq 0$ reduces the charge noise exposure by sweeping between sweet spots. The explanation is simply that the closer the sweet spots are the less time the qubits are exposed to charge noise. Moreover, when the two sweet spots are closer, the derivative $dE_{\rm Larmor}/dF_z$ also becomes smaller. As a result, the exposure to charge noise by sweeping between sweet spots  is minimal when the two sweet spots merge at $\phi=1/2\arcsin(1/3)$ and $\phi=\pi/2-1/2\arcsin(1/3)$. This implies that Protocol 2 is particularly robust against the charge noise exposure during the adiabatic sweep between sweet spots, since the value of $\phi$ for this protocol is close to the divergence of $\frac{T_2^*(\phi)}{T_2^*(0)}$.
\end{widetext}
\end{document}